\documentclass[aps,prd,preprint,superscriptaddress,showpacs,floatfix,
preprintnumbers,nofootinbib, a4paper]{revtex4-1}
\usepackage{graphicx} \usepackage{textcomp}
\usepackage{amsmath,amssymb} \usepackage{enumerate}
\usepackage{gensymb}
\newcommand{\mathsym}[1]{{}}

\newcommand{\ba}{\begin{array}} \newcommand{\ea}{\end{array}}
\newcommand{\be}{\begin{equation}} \newcommand{\ee}{\end{equation}}
\newcommand{\beqa}{\begin{eqnarray}}
\newcommand{\eeqa}{\end{eqnarray}} \def\321{$SU(3)\times SU(2)\times
U(1)$}

\begin{document} \vspace*{1cm}
%ga2ysai
 \title{Perturbed flavour symmetries and predictions of CP violating phase $\delta$} \bigskip
 \author{Anjan S.
Joshipura} \email{anjan@prl.res.in} \affiliation{Physical Research
Laboratory, Navarangpura, Ahmedabad 380 009, India.} 
%\author{Ketan M.
%Patel} \email{ketan.patel@pd.infn.it} \affiliation{Istituto Nazionale
%Fisica Nucleare, Sezione di Padova, I-35131 Padova, Italy.}

%\pacs{11.30.Hv, 14.60.Pq, 11.30.Er}
%--------------------------------------------------------------

\begin{abstract} \vspace*{0.2cm}
{It is known that the imposition of a class of residual $Z_2\times Z_2$
symmetries on
the neutrino mass matrix $M_\nu$ and a residual symmetry $Z_n$ ($n\geq 3$)
on the Hermitian combination $M_lM_l^\dagger$ of the charged lepton mass
matrix leads to a universal prediction of vanishing Dirac CP phase $\delta$
if these symmetries are embedded in $\Delta(6 n^2)$ groups 
and if the leptonic doublets transform as a 3 dimensional irreducible
representation of the group. The Majorana phases remain arbitrary but they
can also be
determined in $\Delta(6 n^2)$ by imposing generalized CP symmetry (GenCP)
consistent with the $\Delta(6 n^2)$ group. We investigate the effects of
adding general perturbations on these predictions assuming that
perturbations break the $Z_2\times Z_2$ symmetry completely but preserve
GenCP. It is found that if the residual symmetries predict the
tri-bimaximal mixing (TBM) among leptons and specific CP conserving values
for the Majorana phases 
then addition of the above perturbations always lead to a neutrino
mass matrix invariant under the $\mu$-$\tau$ reflection symmetry  in the
flavour basis with the
result that perturbations turn the vanishing $\delta$ into maximal value
$\pm \frac{\pi}{2}.$ One gets non-vanishing but generally large $\delta$ if
the predicted zeroeth order mixing deviates from TBM and/or the predicted
Majorana phases are non-trivial. We systematically investigate effects of
perturbations in such situations and work out the predicted $\delta$ for
four of the lowest $\Delta(6 n^2)$ groups with $n=2,4,6,8.$} 
\end{abstract}
\maketitle
\section{Introduction}\label{intro}
Flavour symmetries provide an attractive  framework for the theoretical
understanding  of the leptonic mixing angles and phases. In particular, 
discrete symmetries have been found to be leading to definitive and viable
predictions of these parameters \cite{Lam:2007qc,Lam:2008rs,Lam:2008sh,Lam:2012ga,Lam:2011ag} and these are exhaustively studied in a 
number of papers \cite{Toorop:2011jn,deAdelhartToorop:2011re,Hernandez:2012ra,Hernandez:2012sk,
Holthausen:2012wt,Parattu:2010cy,Fonseca:2014koa,Ge:2011ih}, see
\cite{Altarelli:2010gt,Altarelli:2012ss,Smirnov:2011jv,King:2013eh}.
for
early reviews. It is found that a very large class of discrete subgroups
of $U(3)$   predict vanishing Dirac CP phase $\delta$ \cite{King:2013vna,Joshipura:2016quv}. 
This prediction follows from the following assumptions: ($i$) Neutrino mass
matrix $M_\nu$ is invariant under a $ Z_2\times Z_2\equiv S_\nu$ symmetry
($ii$) The charged lepton mass matrix $M_lM_l^\dagger$ is invariant under a
$ Z_n\equiv T_l$, ($n\geq 3$) group 
($iii$) $S_\nu$ and $T_l$ are contained in a discrete subgroup (DSG) 
$G_f$ of $U(3)$
and ($iv$) three generations of  leptons transform as a 3 dimensional
irreducible representation of $G_f$. Given these assumptions, almost all
the  DSG of $SU(3)$ and many DSG of $U(3)$ taken as $G_f$
\cite{Joshipura:2016quv}  lead
to the prediction $\delta=0.$ In this sense, the prediction $\delta=0$ may
be regarded as a universal prediction following from the above assumptions.
These assumptions  do not fix the neutrino Majorana phases. One can predict
these phases  by  combining   flavour and CP symmetry
\cite{Feruglio:2012cw,Chen:2014wxa,Li:2016ppt,
Hagedorn:2014wha,Ding:2014ora,Holthausen:2012dk} , see \cite{Yao:2016zev} for a detailed list
of references, . Recent
reviews are given in \cite{Petcov:2017ggy,Hagedorn:2017icd}. These
predictions are
explored in details for the $\Delta(6 n^2)$ groups \cite{King:2014rwa} and it is found that the use
of these groups as $G_f$ can lead to non-trivial Majorana phases.

The above  prediction of vanishing $\delta$  appears to be 
at variance with the present experimental trend. The latest results from
the 
NO$\nu$A experiment \cite{Adamson:2017qqn,Adamson:2017gxd} gives
$\sin^2\theta_{23}=0.404 ,\delta=1.48 \pi$ in
case of the normal ordering.
Similarly, recent results from  T2K experiment
\cite{Abe:2017uxa,Abe:2017bay} involving both the neutrino
and anti-neutrino runs gives  $\sin\delta=1.43 \pi.$ A global analysis
of neutrino oscillation data give $\delta=1.40 \pi$ with the 1$\sigma$ range
$(1.20-1.71) \pi$ and  disfavours  the maximal atmospheric  mixing at
$\Delta\chi^2=6.0$ \cite{deSalas:2017kay}, see
\cite{Esteban:2016qun,Capozzi:2016rtj,Gonzalez-Garcia:2014bfa} for other
recent fits. 

The absence of the Dirac CP violation predicted in the above theoretical
framework can change significantly 
in the presence of even small perturbations which would arise from the
breaking of the flavour symmetry.
We wish to systematically analyze here  effects of perturbations to
$M_\nu$
on the prediction  $\delta=0$. We take the
zeroeth order residual symmetry of $M_\nu$  as $S_\nu^{CP}\equiv Z_2\times
Z_2\times H_\nu^{CP}$ as has been done in the general analysis presented
e.g.  in \cite{King:2014rwa}. $H_\nu^{CP}$ here denotes generalized CP (GenCP) transformation commuting with $Z_2\times Z_2.$ We however  allow for the most general perturbations to it  which
break $Z_2\times Z_2$ symmetry completely  but preserves GenCP  contained in
$S_\nu^{CP}$. It is possible to study effect of such perturbations in a
model independent manner as we shall show.
These perturbations have dramatic effect on the prediction of the  Dirac CP
phase. We show that  if $Z_2\times Z_2$ symmetry predicts 
tri-bimaximal mixing  (TBM) pattern and if  the group theoretically determined Majorana phases\footnote
{Following \cite{Patrignani:2016xqp}, we denote the Majorana phase matrix
on the RHS of
$U_{PMNS}$ as ${\rm
diag.}(1,e^{\frac{\alpha_{21}}{2}},e^{\frac{\alpha_{31}}{2}})$}
$\alpha_{21},\alpha_{31}$ are predicted to have CP conserving values
$0,\pi$ respectively at the leading order then
switching on the generalized CP invariant perturbations lead to the
prediction of the maximal CP phases and the maximal atmospheric mixing
angle for a 
class of GenCP symmetry. Even when the zeroeth order mixing matrix  $U_{0}$
does not have
the TBM form or the Majorana
phases are non-trivial, one still  gets quite large Dirac  CP phases in the
presence of perturbations. The  predicted
Dirac phases in several cases are characteristic of the underlying residual
symmetries rather than the values of the perturbation parameters. We
numerically derive such predictions for the residual symmetries contained in
subgroups of first four  groups\footnote{Only groups with even $n$ are
relevant here since $\Delta(6 n^2)$ with odd $n$ do not contain
$Z_2\times Z_2$ groups as subgroups.} in the $\Delta(6 n^2)$ series with
$n=2,4,6,8.$

We first review basic consequences of imposing residual symmetries in the
next section. Then we derive  
general form of the leptonic mixing matrix in the presence of the GenCP
invariant perturbations in section \ref{gencp}. Section \ref{sec6nsquare}  discusses possible
residual symmetries in the context of the $\Delta(6 n^2)$ groups followed
in section \ref{theorem} by a discussion of conditions under which one
obtains the maximal  Dirac phase. Explicit form of the perturbations in
$\Delta(6n^2)$ group is presented in section \ref{explicitmnu}. This is
followed by discussion of numerical results in section \ref{numerical}. The
last section gives a summary. 
\section{Formalism}\label{formalism}
We briefly review here consequences of imposing residual symmetries on the leptonic mass matrices.
The leading order Majorana mass matrix for the neutrinos in some symmetry
basis is defined as $M_{0\nu}$  and the charged lepton mass matrix as $M_l$
They are assumed to satisfy symmetry relations
\cite{Altarelli:2010gt,Altarelli:2012ss,Smirnov:2011jv,King:2013eh}
\beqa \label{smnu0}
S_{1\nu,2\nu}^T M_{0\nu}S_{1\nu,2\nu}&=& M_{0\nu}~,\nonumber \\
T_l^\dagger M_lM_l^\dagger T_l&=&M_lM_l^\dagger~.\eeqa
$S_{1\nu}$ and $S_{2\nu}$ are $3\times 3$ unitary matrices generating the
group $Z_2\times Z_2$ and $T_l$ generates a $Z_n,n\geq 3$. Let $U_{S_\nu}$
be a unitary matrix which 
diagonalizes $S_{1\nu}$ and $S_{2\nu}$ simultaneously. Explicitly,
\be \label{s1s2dia}
U_{S_\nu}^\dagger S_{1\nu}U_{S_\nu}=
diag.(-1,-1,1)~~,~~U_{S_\nu}^\dagger S_{2\nu}U_{S_{\nu}}=
diag.(1,-1,-1)~,\ee
$U_{S_\nu}$ is arbitrary upto a multiplication by a diagonal phase matrix
from right. This arbitrariness can be fixed by imposing CP as an
additional symmetry and taking $S_\nu^{CP}\equiv Z_2\times Z_2\times 
H^{CP}_\nu$ as the complete residual  symmetry.  The action of GenCP on the
neutrino
triplet is represented by a $3\times 3$ symmetric unitary matrix $X_\nu$.
Requiring that action of each of the $Z_2$ separately commutes with GenCP
operation  imposes the constraints \cite{Feruglio:2012cw}:
\be\label{cps}
X_\nu S_{1\nu,2\nu}^*X_\nu^\dagger=S_{1\nu,2\nu}~.\ee
Invariance of the neutrino mass term under GenCP requires
\be \label{cpmnu0}
X_\nu^T M_{0\nu}X_\nu=M_{0\nu}^*~.\ee
Eqs.(\ref{smnu0}-\ref{cpmnu0}) are sufficient to completely determine the
neutrino mixing matrix.
If we define $\hat{X_\nu}\equiv  U_{S_\nu}^\dagger
X_\nu U_{S_\nu}^*$, then eqs.(\ref{s1s2dia},\ref{cps}) imply 
\be\label{pnu}
\hat{X_\nu}\equiv
P_\nu^2 ~,\ee
where $P_\nu$ is a diagonal phase matrix.  This then implies
\be \label{v0nu}
X_\nu=V_{0\nu}V_{0\nu}^T~,V_{0\nu}\equiv U_{S_\nu}P_\nu~.\ee
$V_{0\nu}$ in this way gets determined from the structure of $S_{1\nu,2\nu}$
and $X_\nu$. Eq.(\ref{smnu0}) implies that $M_{0\nu}$ is diagonalized by
$V_{0\nu}$:
\be \label{diamnu0}
V_{0\nu}^T M_{0\nu}V_{0\nu}=D_0\equiv {\rm diag,}(m_1,m_2,.m_3)~.\ee
GenCP invariance of $M_{0\nu}$, eq.(\ref{cpmnu0}) then implies that $D_0$
defined above is a real matrix. Thus
$V_{0\nu}$ diagonalizes $M_{0\nu}$ with real (not necessarily positive) 
eigenvalues and can be taken as
the neutrino mixing matrix at the leading order.
 
The complete mixing matrix  at the leading order is given by
\be \label{u0pmns}
U_{0}\equiv U_l^\dagger U_{S_\nu}P_\nu\equiv U_{S}P_\nu
~, \ee  
where $U_l$ is a matrix that diagonalizes $M_lM_l^\dagger.$ It is determined
up to overall phases by 
its
residual symmetry $Z_n$ if $n>2.$
\section{Generalized CP invariant perturbations and mixing
matrix}\label{gencp}
We now discuss the effect of GenCP invariant perturbations on the structure of
the mixing matrix, eq.(\ref{u0pmns}). Assume that  the neutrino mass matrix
has the form
\be \label{mnu}
M_\nu=M_{0\nu}+\delta M_\nu~.\ee
$\delta M_\nu$ is a perturbation matrix which would arise from the
$Z_2\times Z_2$ symmetry breaking in models.
We assume that GenCP symmetry is not broken at this stage and $\delta M_\nu$
thus satisfies
\be 
\label{xdelmu}
X_\nu^T\delta M_\nu X_\nu=\delta M_\nu^*~.\ee
This assumption leads to the following  general structure of the mixing
matrix $U_{sym}:$
\be \label{upmns}
U_{PMNS}\sim U_{sym}=U_SP_\nu OK~.\ee
Here $U_S\equiv U_l^\dagger U_{S_\nu}$, see eq.(\ref{u0pmns}) is a matrix
determined by the residual symmetries.
$P_\nu$ is a diagonal phase matrix determined by $X_\nu$ and $O$ is a real
orthogonal matrix resulting from the perturbations. $K$ is a diagonal phase
matrix with elements $\pm 1,\pm i$ which is used to make the eigenvalues of
$M_\nu$ positive. The residual symmetry cannot predict the order of
the leptonic masses and hence orders of the rows and columns in $U_{sym}.$
The correspondence between $U_{PMNS}$ shown on the left and $U_{sym}$ has
to be decided on the phenomenological grounds in these symmetry based
approaches.
 
Eq.(\ref{upmns}) follows in a straightforward way. We first re-express
eq.(\ref{mnu}) in the basis with a
diagonal $M_{0\nu}$ by defining $M_\nu^\prime \equiv V_{0\nu}^T M_\nu
V_{0\nu}$.
This gives
\be \label{mnuprime}
M_\nu^\prime=D_0+\delta M_\nu^\prime ~,\ee
where $D_0$ is defined in eq.(\ref{diamnu0}) and $\delta M_\nu^\prime\equiv
V_{0\nu}^T \delta M_\nu V_{0\nu}$. Then use of eq.(\ref{xdelmu}) together with
$X_\nu=V_{0\nu} V_{0\nu}^T$ implies
$$\delta M_\nu^\prime=\delta M_\nu^{\prime *}~.$$
Since $D_0$ is also real it follows that $M_\nu^\prime$ is a real symmetric
matrix which can be diagonalized by an orthogonal matrix $O$. Thus neutrino
mass matrix $M_{\nu}$ is diagonalized by $V_\nu O=U_{S_\nu}P_\nu O$ and one gets
$U_{PMNS}$ as given in eq.(\ref{upmns}).

It turns out that  $U_{S}$ in eq.(\ref{upmns})  can be made real  by absorbing
all its phases in $P_\nu$ or in redefining the charged lepton fields when
the residual symmetries are embedded in $\Delta(6 n^2)$ groups. With $O$
also  real, the only source of CP violation in
eq.(\ref{upmns}) is the (appropriately redefined) phase matrix $P_\nu.$
This corresponds to only Majorana CP violation at the leading order but
$P_\nu$ plays non-trivial role and generates Dirac CP violation when $O$ is
present. Remarkably, the non-trivial Dirac phase can result
even in the CP conserving situation at the tree level corresponding to
trivial  Majorana phases as we discuss now.
\section{$\Delta (6 n^2)$ symmetry} \label{sec6nsquare}
We first  outline possible  choices of Klein groups and the CP symmetries
consistent with them in the context of the $\Delta (6 n^2)$ groups
\cite{King:2013vna,King:2014rwa} to set
our notations. This symmetry group is generated by four elements  
$a,b,c,d$ satisfying
$$a^3=b^2=(ab)^2=c^n=d^n~.$$
These elements are represented in one of the three-dimensional
representation of $\Delta(6 n^2)$ ($3^1_2$ in the notation of \cite{King:2013vna}) as
\be \label{generators}
A=\left(\ba{ccc} 0&1&0\\0&0&1\\1&0&0\\ \ea \right)~;~
B=-\left(\ba{ccc} 0&0&1\\0&1&0\\1&0&0\\ \ea \right)~;~
C=\left(\ba{ccc}\eta&0&0\\0&\eta^*&0\\0&0&1\\\ea \right)~;~
D=\left(\ba{ccc}1&0&0\\0&\eta&0\\0&0&\eta^*\\\ea \right)~,\ee
where $\eta=e^{\frac{2\pi i}{n}}$ and we denote the 3-dimensional
representation of elements by the corresponding capital letters. A set of
Klein groups within $\Delta(6 n^2)$ is generated from
\be \label{s1nus2nu}
S_{1\nu}\equiv BC^{\gamma_\nu}D^{\gamma_{\nu}}=-\left(\ba{ccc} 
0&0&\eta^{-\gamma_\nu}\\
0&1&0\\
\eta^{\gamma_\nu}&0&0\\  \ea
\right)~;~
S_{2\nu}\equiv BC^{\gamma_\nu+\frac{n}{2}}D^{\gamma_{\nu}+\frac{n}{2}}
=-\left(\ba { ccc } 
0&0&-\eta^{-\gamma_\nu}\\
0&1&0\\
-\eta^{\gamma_\nu}&0&0\\  \ea
\right)~,\ee
where $\gamma_\nu=0,1,2....n-1.$
These two alongwith their products and squares form a set of  Klein group
$Z_2\times Z_2$. One could obtain two other sets 
from the cyclic permutations of these. Mixing patterns predicted in
all three cases are equivalent and we will
specifically use eq.(\ref{s1nus2nu}) as neutrino symmetries. The other possible Klein groups
within $\Delta(6 n^2)$ consist of diagonal generators in the chosen basis. They lead to 
democratic mixing at the leading order when $T_l=Z_n$ and thus predict
$\sin^2\theta_{13}=\frac{1}{3}$ which is far from the actual value.  We
shall therefore  not consider them.

The minimal requirements on CP symmetry $X$ \cite{Feruglio:2012cw} is that it should satisfy
eq.(\ref{cps}) which ensures that the residual symmetry
$ H^{CP}_\nu$ commutes with $S_{1\nu},S_{2\nu}.$ If
this
symmetry is to be embedded in a flavour group $G_f$ then there are further
requirements on $X$ for the consistent definition of $G_f$ and $CP$
\cite{Holthausen:2012dk,Chen:2014tpa}. These
are studied at length in general situations and in the context of the
$\Delta(6 n^2)$ groups \cite{King:2014rwa}. One basically requires that $X_r$ for every
representation $\rho_r$   of $G_f$ should satisfy
$$X_r \rho_r(g)^*X_r^\dagger=\rho_r(g')~,$$
where  $g$ and  $g'$ are elements of $G_f$ and the above equation should
remain true for every $g\in G_f.$ 
It has been argued \cite{King:2014rwa} in the context of $G_f=\Delta(6 n^2)$ that $X$ should be
an element of the group satisfying $X^T =X,XX^*=1$ upto an overall phase.
Two possible sets of  $X$ within $\Delta(6 n^2)$ are given in three
dimensional representation as
\be \label{x1x2}
X_{1\nu}\equiv C^xD^{-x-2\gamma_\nu}=\left( \ba{ccc}
\eta^x&0&0\\
0&\eta^{-2 x-
2\gamma_\nu}&0\\
0&0&\eta^{x+2\gamma_\nu}\\ \ea \right)
~~,~~X_{2\nu}\equiv C^xD^{-x}B=-\left( \ba{ccc}
0&0&\eta^x\\
0&\eta^{-2 x}&0\\
\eta^{x}&0&0\\ \ea \right)~, \ee
where $x=0,1....n-1$. Both of these satisfy the required eq.(\ref{cps}) for
$S_{1\nu,2\nu}$ given by eq.(\ref{s1nus2nu}).
We shall use these two choices of $X$ and study consequences of
imposing these on total $M_\nu$, eq.(\ref{mnu}).
 
Common matrix diagonalizing $S_{1\nu},S_{2\nu}$ is given by
\be \label{usnu}
U_{S_\nu}=\frac{1}{\sqrt{2}}\left(
\ba{ccc}
\eta^{-\frac{\gamma_{\nu}}{2}}&0&\eta^{-\frac{\gamma_\nu}{2}}\\
0&\sqrt{2}&0\\
\eta^{\frac{\gamma_\nu}{2}}&0&-\eta^{\frac{\gamma_\nu}{2}}\\
\ea \right)~.\ee
The neutrino mixing matrix $V_{0\nu}$, eq.(\ref{v0nu}) is obtained from the
above $U_{S_\nu}$ by multiplying
it with the phase matrix $P_\nu$ which is determined by $X_{1\nu,2\nu}$ as
given in
eq.(\ref{pnu}). Denoting $P_\nu$ in these cases by $P_{1\nu,2\nu}$, we have 
\be \label{p0nu}
P_{1\nu}=\left(
\ba{ccc}
p_1&0&0\\
0&p_1^{-2}&0\\
0&0&p_1\\ \ea \right)~~;~~ P_{2\nu}=\left(
\ba{ccc}
p_2&0&0\\
0&p_2^{-2}&0\\
0&0&-i p_2\\ \ea \right)~,\ee
with $p_1=e^{\frac{i\pi (x+\gamma_\nu)}{n}}$ 
and  $p_2=e^{\frac{i\pi x}{n}}.$ 

The matrix $U_l$ is determined by the symmetry $T_l$ of $M_lM_l^\dagger.$
This symmetry group is chosen as a set of $Z_n$ groups defined in the three
dimensional representation as:
\be  \label{tl}
T_l\equiv C^{l_1}D^{l_1+l_2}A=\left(\ba{ccc}
0&\eta^{l_1}&0\\
0&0&\eta^{l_2}\\
\eta^{-(l_1+l_2)}&0&0\\
\ea \right)~, \ee
with $l_1,l_2=0,1...n-1.$ Other possible $Z_n$ sub-groups of $\Delta(6 n^2)$  are diagonal or block diagonal
(analogous to $S_{1\nu,2\nu}$). The only zeroeth order viable pattern of
mixing predicted in these cases \cite{Joshipura:2016quv} correspond to the democratic or  bi-maximal mixing. Since
the solar mixing angle at the zeroeth order considerably deviates
from its actual value in this mixing pattern, we will omit such groups from the discussion
and work with the set of $Z_n$ define by eq.(\ref{tl}). 
$T_l$ is diagonalized by
\be \label{ul}
U_l\equiv P_l U_\omega =\frac{1}{\sqrt{3}}\left(
\ba{ccc}
\eta^{l_1}&0&0\\
0&1&0\\
0&0&\eta^{-l_2}\\ \ea\right)
\left(\ba{ccc}
1&1&1\\
1&\omega&\omega^2\\
1&\omega^2&\omega\\
\ea \right)~ \ee
with $\omega=e^{\frac{2 i\pi}{3}}.$
The complete mixing matrix following from the above determined $V_{0\nu}$,
$U_l$ and eq.(\ref{upmns}) can be written as:
\be \label{upmns6nsquare}
U_{sym}=p_a\eta^{\frac{l_2-l1}{2}}{\rm
diag.}(1,\omega^2,\omega)\frac{1}{\sqrt{3}}\left(\ba{ccc}
\sqrt{2} c_1&1&-\sqrt{2} i s_1\\
\sqrt{2} c_2&1&-\sqrt{2} i s_2\\
\sqrt{2} c_3&1&-\sqrt{2} i s_3\\ \ea \right) {\rm.
diag.}(1,p_a^{-3}\eta^{\frac{l_1-l_2}{2}},\epsilon_a)O~,\ee
We distinguish two symmetries $X_{a\nu}$ by label $a=1,2.$ $p_a$
arising from $P_{a\nu}$ are defined below eq.(\ref{p0nu}) and
$\epsilon_1=1,\epsilon_2=-i$.
$c_i=\cos \theta_i,s_i=\sin
\theta_i$ with 
\be \label{thetai}
\theta_1= \frac{\pi \gamma}{n}\equiv\frac{\pi
(l_1+l_2+\gamma_{\nu})}{n}~,~ \theta_2=\theta_1+\frac{4
\pi}{3}~,~\theta_{3}=\theta_1+\frac{2 \pi}{3}~.\ee
$U_{sym}$ represent the PMNS mixing matrix as given by the underlying symmetry.
The phase matrix appearing in the LHS  of the above equation and an
overall phase can be removed
by redefining the charged lepton phases and will be neglected further. 
At the leading order, $O$ is an identity matrix and the above $U_{sym}$
leads to vanishing Dirac phase as already noted in \cite{King:2014rwa} .
Non-trivial $O$
arises
in the presence of perturbations. We shall discuss possible nature 
of perturbations in the next section. Here we present an important
consequence of eq.(\ref{upmns6nsquare}) which purely follows from symmetry
rather than details of the perturbations.
\subsection{From conserved to maximal CP}\label{theorem}
We now show that the perturbations can completely change the
prediction of
vanishing Dirac CP phase which can even take the maximal value
$\sin\delta=\pm
1$. Interestingly, the occurrence of the maximal phase is intimately
tied to the absence of CP violation in the leading order PMNS matrix.
CP is conserved at the leading order if the relevant Majorana
phases
$\alpha_{21},\alpha_{31}$ are 0 or $\pi.$ Group theoretically determined 
phase $\alpha_{31}$ from 
eq.(\ref{upmns6nsquare})   is  $\pi(0)$
in case of the symmetries $X_{1\nu} (X_{2\nu})$ independent of the values of
$x,\gamma.$
$\alpha_{21}=0,\pi$ when $p_a^{-3}\eta^{\frac{l_1-l_2}{2} }=1,i$ respectively.
In this case, the Dirac as well Majorana CP violation is absent at the
leading order. Turning on perturbations can change this completely.
Specifically, we show that the following result holds:

\noindent If the underlying $Z_2\times Z_2$ symmetry leads to the
tri-bimaximal mixing
pattern at the leading order
and if the CP violating Majorana phase $\alpha_{21}$ as predicted by CP
symmetry
$X_{1\nu}$ assumes value $0$ or $\pi$   then an   arbitrary GenCP
invariant perturbations correcting for the original tri-bimaximal mixing
lead to a theory with maximal CP phase and maximal atmospheric mixing angle.

The proof of the above  follows in a straightforward manner . Consider the
elements $|U_{\alpha i}|^2$ for $\alpha=\mu,\tau$:
\beqa\label{modulie}
|U_{\mu i}|^2&=&\frac{1}{3}| \sqrt{2} c_2 O_{1
i}+p_1^{-3}\eta^{\frac{l_1-l_2}{2} }O_{2 i}- i
\sqrt{2}  s_2 O_{3 i}|^2\nonumber ~,\\
|U_{\tau i}|^2&=&\frac{1}{3}|\sqrt{2}c_3 O_{1
i}+p_1^{-3}\eta^{\frac{l_1-l_2}{2} }O_{2 i}-i
\sqrt{2} s_3 O_{3 i}|^2~.\eeqa
We have chosen here  specific order
in which the second and the third row of $U_{sym}$  are taken to be
associated
with the $\mu$
and $\tau$ flavours respectively. The same result would follow for the other
choices but with a different value for the angle $\frac{\pi \gamma}{n}$.
Simultaneous occurrence of the maximal CP phase and $\theta_{23}$ is termed
as $\mu$-$\tau$ reflection symmetry \cite{Harrison:2002et} and is obtained
from the following
relation:
\be \label{mtr}
 |U_{\mu i}|=|U_{\tau i}| ~.\ee
If we do not want any fine tunning then a  prerequisite to obtain the above
relation with perturbations is that the zeroeth order mixing matrix as
implied from the $Z_2\times Z_2$ symmetry also satisfies this relation. This
requires either
$(a)~ s_2=-s_3$ or $(b)~s_2=s_3$ in eq.(\ref{upmns6nsquare}). 
These cases lead to  the tri-bimaximal pattern at the
leading order since they  imply $\theta=0$ for case $(a)$ and $\theta=\pm \frac{\pi}{2}$ for case ($b$). The third column of $U_0$ has TBM form in case ($a)$ while for case ($b$)
the first column of $U_0$ has the TBM  form $(0,-\frac{1}{\sqrt{2}},\frac{1}{\sqrt{2}})^T$ and one gets
phenomenologically consistent picture in this case by identifying the first
column with the heavier (lighter) mass eigenstate for the normal (inverted)
hierarchy. Ii is seen that  eq.(\ref{modulie}) leads to
eq.(\ref{mtr}) for an arbitrary $O$ if $p_1^{-3}\eta^{\frac{l_1-l_2}{2} }$ takes the value $\pm 1$ for case ($a$) and $\pm i$ for ($b$)
which as discussed before is equivalent to
requiring the Majorana phase $\alpha_{21}=0,\pi.$
The $\mu$-$\tau$ reflection symmetry is known to lead to
$s_{23}=\frac{\pi}{4}$ and $s_{13} \cos \delta=0$ \cite{Grimus:2003yn}.
Thus perturbations $O$
correcting for the vanishing $s_{13}^2$  invariably lead to the maximal CP violation. 
We note that 
\begin{itemize}
\item While the phase restriction
$p_1^{-3}\eta^{\frac{l_1-l_2}{2} }=\pm 1,i$ is necessary to
obtain the
maximal CP phase for the most general perturbations, there exists a special
class of perturbations for which the maximal CP phase follows independent of
of this. This happens when
the matrix $O$ is a pure rotation in the $13$-plane. In this case, the phase
matrix appearing on the right hand side of eq.(\ref{upmns6nsquare})  commutes
with $O$ and the Dirac CP phase and the mixing
angles become independent of the Majorana phases. In this case,
the tribimaximal mixing matrix automatically leads to the $\mu$-$\tau$
reflection symmetry in the presence of perturbations. The neutrino mass matrix in this situation is
invariant
under a $Z_2\times H_\nu^{CP}.$ This special case is already discussed in
the literature \cite{
Feruglio:2012cw,Holthausen:2012dk,Chen:2014wxa,Li:2016ppt,
Hagedorn:2014wha,Ding:2014ora,Yao:2016zev}. However as shown here, the
occurrence of the maximal phase
is
more general and one does not need to assume any unbroken $Z_2$ symmetry in
order to
get the above result which holds for arbitrary GenCP invariant  perturbations.\\
\item The lowest member of the $\Delta(6 n^2)$ groups namely, $S_4$ contains the
residual symmetry needed for the prediction
$p_1^{-3}\eta^{\frac{l_2-l_1}{2}}=\pm 1$. The other residual symmetries
predicting $p_1^{-1}\eta^{\frac{l_1-l_2}{6}}=\pm (\omega,\omega^2)$ and
hence  $p_1^{-3}\eta^{\frac{l_1-l_2}{2}}=\pm 1,\pm i$ arise
in $\Delta(6 n^2)$ series with $n=6 k,k=1,2...$ with the lowest order group
in the series  being $\Delta(216)$.
\item The maximality of the phase essentially arises from the factor $i$
present in the third column of eq.(\ref{upmns6nsquare}) when the predicted
phase $p_1^{-3}\eta^{\frac{l_1-l_2}{2} }$ is $\pm 1$.  The relative factor
of $i$ 
is essentially produced   by the structure of the underlying $Z_2\times Z_2$
symmetry. Such factors  do not play any role in the  CP violation at the
leading order and the $U_{PMNS}$ matrix therefore is taken sometimes to be real in the
literature. Here it plays an important role in giving CP violation when 
perturbations are introduced.
\item The genesis of the the $\mu$-$\tau$ reflection symmetry appearing here
can be easily understood on general grounds. 
It is known that this symmetry can be obtained if the neutrino mass matrix
in the flavour basis corresponding to the diagonal charged lepton mass
matrix  satisfies \cite{Grimus:2003yn}
\be \label{mtcp}
S^T M_{\nu f} S=M_{\nu f}^*~,\ee
where $S$ denotes a $Z_2$ symmetry which interchanges $\mu$ and $\tau.$
It is not difficult to see that if the conditions outlined above are satisfied then
one indeed gets eq.(\ref{mtcp}) as an effective symmetry of $M_{\nu f}.$
This can be seen by expressing $X_{1\nu}$ in the flavour basis.
It can be written in this basis as 
$$\tilde{X}_{1\nu}= U_l^\dagger X_{1\nu} U_l^*$$
where $U_l$ diagonalizes $T_l$ and hence $M_lM_l^\dagger$ is given by eq.(\ref{ul}). One then finds 
$$ \tilde{X}_{1\nu}=U_\omega^\dagger P_l^*X_{1\nu}P_l^*
U_\omega^*=U_\omega^\dagger {\rm diag.}(q_1,q_2,q_3)U_\omega^*~,$$
where 
$$ q_1=\eta^{x-2 l_1}~~,~~q_2=\eta^{-2 x-2
\gamma_\nu}~~,~~q_3=\eta^{x+2\gamma_\nu+2 l_2}~.$$
The zeroeth order mixing would be TBM for $\gamma=(l_1+l_2+\gamma_\nu)=0,n,2
n...$ and the triviality of $\alpha_{21}$ would follow if $l_1-l_2-3
(x+\gamma_\nu)=0,n, 2n.... $. It is easy to show that if both these
conditions are satisfied then all $q_i$ are proportional to a complex phase
in general and $\tilde{X}_{1\nu}\approx U_\omega^\dagger U_\omega^*= S$
and
GenCP condition  (\ref{cps}) is equivalent to eq.(\ref{mtcp}). 
\item Two GenCP symmetries $X_{1\nu,2\nu}$ differ in their prediction of
$\alpha_{31}$ at the leading order, see eq.(\ref{p0nu}). This leads to
different predictions in these cases. Specifically, eq.(\ref{modulie}) now
becomes
\beqa\label{modulie2}
|U_{\mu i}|^2&=&\frac{1}{3}| \sqrt{2} c_2 O_{1
i}+p_2^{-3}\eta^{\frac{l_1-l_2}{2} }O_{2 i}-
\sqrt{2}  s_2 O_{3 i}|^2\nonumber ~,\\
|U_{\tau i}|^2&=&\frac{1}{3}|\sqrt{2}c_3 O_{1
i}+p_2^{-3}\eta^{\frac{l_1-l_2}{2} }O_{2 i}-
\sqrt{2} s_3 O_{3 i}|^2~.\eeqa
This equation does not lead to the $\mu$-$\tau$ reflection symmetry
even if $p_2^{-3}\eta^{\frac{l_1-l_2}{2} }=\pm 1$ and zeroeth order mixing
is TBM, i.e. $s_2=-s_3$.
\item The leading order prediction of the $Z_2\times Z_2$ symmetry may not
be TBM mixing or the phase $\alpha_{21}$ may not be trivial. In either case,
one can
get   large but non-maximal phase  even in the presence of very small
perturbations.  We will study such cases numerically. 
\end{itemize}
\section{$M_\nu$ with perturbed $Z_2\times Z_2\times
H_{\nu}^{CP}$}\label{explicitmnu}
In this section, we construct  a general
neutrino mass matrix $M_\nu$ with the broken   $Z_2\times Z_2$ but intact 
GenCP for the $\Delta(6 n^2)$ groups and work out approximate expressions for the matrix
which diagonalizes it. We explicitly discuss only the case of $X_{1\nu}$ in
view of the fact that it can lead to near maximal $\delta$. The other
symmetry $X_{2\nu}$ can be analogously discussed.
The $U_{\rm PMNS}$ matrix in eq.(\ref{upmns6nsquare}) contains three unknown
mixing angles of the matrix $O$. One could analyze general predictions in
terms of these angles. Here we adopt an alternative parameterization.
Explicit representation of $S_{1\nu},S_{2\nu}$, eq.(\ref{s1nus2nu}) and CP
operators $X_{1\nu}$ can be used to construct the $Z_2\times
Z_2\times
H_\nu^{CP}$ symmetric leading order neutrino mass matrix  $M_{0\nu}.$
Explicitly.
\be \label{m0nudelta}
M_{0\nu}\equiv V_{0\nu}^*{\rm diag.}(m_1,m_2,m_3)V_{0\nu}^\dagger=\left(
\ba{ccc}
\frac{1}{2}(m_1+m_3)\eta^{-x}&0&\frac{1}{2}(m_1-m_3)\eta^{-\gamma_\nu-x}
\\
0&m_2\eta^{2 x+2\gamma_\nu}&0\\
\frac{1}{2}(m_1-m_3)\eta^{-\gamma_\nu-x}&0&\frac{1}{2}(m_1+m_3)\eta^{
-2\gamma_\nu-x }\\
 \ea \right) ~.\ee
The most general    perturbation matrix $\delta M_\nu$  satisfying
eq.(\ref{xdelmu}) can be written, after appropriate redefinition of the
unperturbed masses in eq.(\ref{m0nudelta})  as:
\be \label{delmnudelta}
\delta M_\nu=m_3\left(
\ba{ccc}
-\epsilon_{33}\eta^{-x}&\epsilon_{12}\eta^{\gamma_\nu+\frac{x}{2}}&0\\
\epsilon_{12}\eta^{\gamma_\nu+\frac{x}{2}}&0&\epsilon_{23}\eta^{\frac{x}{2}}
\\
0&\epsilon_{23}\eta^{x/2}&\epsilon_{33}\eta^{-x-2\gamma_\nu}\\
 \ea \right) ~,\ee
where $m_3$ is the heavier mass in the case of the normal hierarchy. $\delta M_\nu$  is characterized by three real parameters
$\epsilon_{33},\epsilon_{12},\epsilon_{23}$. Eq.(\ref{mnu}) assumes a
simple form when transformed to a basis in which  the unperturbed matrix 
$M_{0\nu}$ is diagonal:
\be \label{mnuhat}
\tilde{M}_\nu\equiv V_{0\nu}^T M_\nu V_{0\nu}
=\left(
\ba{ccc}
m_1&m_3\frac{\epsilon_{12}+\epsilon_{23}}{\sqrt{2}}&-m_3\epsilon_{33}\\
m_3\frac{\epsilon_{12}+\epsilon_{23}}{\sqrt{2}}&m_2&m_3\frac{\epsilon_{12}
-\epsilon_{23}}{\sqrt{2}}\\
-m_3\epsilon_{33}&m_3\frac{\epsilon_{13}-\epsilon_{23}}{\sqrt{2}}&m_3\\
\ea \right) \ee
$\tilde {M_\nu}$ being real is diagonalized by an orthogonal matrix $O$ and
thus $M_\nu$ is diagonalized by $V_\nu=V_{0\nu}O$ leading to the $U_{PMNS}$
matrix as given in eq.(\ref{upmns6nsquare}). 

These perturbations reduce to the known cases in specific limits. When
$\epsilon_{12}=\epsilon_{23}=0$, $O$ is a pure rotation in the $1-3$ plane
with an angle $\phi_{13}$. One of the $Z_2\times Z_2$ symmetry namely, the
one corresponding to the tri-maximal mixing of the second column of the
$U_{0PMNS}$ remains unbroken in this case. A different $Z_2$ remains
unbroken when  $\epsilon_{33}=0,\epsilon_{23}=-\epsilon_{12}$. The first
column of the PMNS matrix coincides with the zeroeth order result in this
case. Finally, the third column of the $PMNS$ matrix remains unaffected when
$\epsilon_{33}=0, \epsilon_{12}=\epsilon_{23}$. In this case, $O$ is a pure
rotation in the $1-2$ plane. Consequences of these  single $Z_2\times
H_{\nu}^{CP}$ have been extensively studied in the so-called semi-direct
approach in a number of works
\cite{Feruglio:2012cw,Holthausen:2012dk,Chen:2014wxa,Li:2016ppt,
Hagedorn:2014wha,Ding:2014ora,Yao:2016zev}.
The matrix $O$ diagonalizing $\hat{M_\nu}$ can be determined perturbatively.
We parameterize $O$ as
\be \label{oparam}
O=R_{23}(\phi_{23}) R_{13}(\phi_{13}) R_{12}(\phi_{12})~,\ee
where the rotation $R_{ij}(\phi_{ij})$ denotes a  rotation in the $ij^{th}$
plane by  an angle $\phi_{ij}.$ $O$ satisfies
\be 
O^T\hat{M_\nu} O={\rm diag.}(m_{1\nu},m_{2\nu},m_{3\nu})~\ee
Mixing angles are approximately  given by
\beqa \label{phiij}
\sin\phi_{23}&\approx& \frac{m_3}{(m_3-m_2)}\epsilon_-
-\frac{m_3^2}{(m_3-m_1)(m_3-m_2)}\epsilon_+\epsilon_{33}~,\nonumber\\
\sin\phi_{13}&\approx& \frac{m_3}{(m_3-m_1)}\epsilon_{33}
-\frac{m_3^2}{(m_3-m_1)(m_3-m_2)}\epsilon_+\epsilon_{-}~,\nonumber\\
\sin\phi_{12}&\approx& -\frac{m_3}{(m_2-m_1)}\epsilon_+
-\frac{m_3^2}{(m_3-m_2)(m_2-m_1)}\epsilon_-\epsilon_{33}~.\eeqa

The above mixing angles diagonalize  $\hat{M_\nu}$ modulo ${\cal O}(\epsilon^3)$ corrections.
The neutrino masses receive corrections only at the second order in
perturbation parameters and are given by:
\beqa \label{masses}
m_{\nu_1}&\approx& m_1-\frac{\epsilon_{33}^2m_3^2}{m_3-m_1}-\frac{\epsilon_+^2m_3^2}{m_2-m_1}~,\nonumber \\
m_{\nu_2}&\approx&
m_2-\frac{\epsilon_{-}^2m_3^2}{m_3-m_2}+\frac{\epsilon_+^2m_3^2}{m_2-m_1}~,
\nonumber \\
m_{\nu_3}&\approx&
m_3\frac{\epsilon_{-}^2m_3^2}{m_3-m_2}+\frac{\epsilon_{33}^2m_3^2}{m_3-m_1}
~.\eeqa

It is seen that the  mixing angles $\phi_{ij}$  not only depend on the
strength of the perturbations $\epsilon_{ij}$ but also on the
relative signs
of the unperturbed masses $m_{i}$. Thus equal (opposite) signs of $m_i$ and
$m_j$
tend to magnify or (suppress) the mixing angle $\phi_{ij}$ particularly for $\phi_{12}$. In this way, the
effect
of a small perturbations can get magnified. This allows one to obtain
correct structure of the final mixing matrix even with very small
perturbations as we will explicitly see in the numerical analysis to be
presented in the next section. 
\section{Numerical results for specific groups} \label{numerical}
We now consider the four lowest groups of the $\Delta(6 n^2)$ series and
explore their predictions for the CP violating phases. We only consider the
symmetry $X_{1\nu}$ which leads to rather large $\delta$ in many cases. 
The structure of the
$U_{0}$ with the imposed residual symmetries is characterized by three
integers $(n,\gamma,x)$ defined earlier. We shall choose specific values of
$\gamma$ for a given $n$ such that  the third column of the  $U_{0}$ or
its cyclic permutations provide a good zeroeth order approximations to the
experimental values and then explore the influence of perturbation for each
of the possible CP symmetries characterized by $x$. From now on, we
specialize to the case with $l_1=l_2=0$ in eq.(\ref{tl}) 
in which case $T_l$ is a $Z_3$ symmetry. Other choices of $l_1,l_2$ give
equivalent results.
The tri-bimaximal mixing is the only such  possibility at the zeroeth order 
for the group $S_4$,  obtained when $\gamma=0$ in eq.(\ref{thetai}).
$\Delta(96)$
allows one more possibility with the  third column
$|U_{3i}|^2=(0.044,0.333,0.622)^T$. This mixing pattern   obtained with
$\gamma=1,n=4$ (or their integer multiples)  \cite{Toorop:2011jn} is known
as the Toorop,Feruglio, Hagedorn (TFH) 
mixing. The
group $\Delta(384)$ contains additional possibility  corresponding to
$\gamma=3$ which leads after permutation of the third  column in
eq.(\ref{upmns6nsquare})  to values
$(0.011,0.419,0.569)^T.$ This can fit the mixing angles $s_{13},s_{23}$
when small perturbations  are considered.  The 
TFH mixing on the other hand requires somewhat larger corrections to
$s_{13}^2$ and we do not consider it here.

Before discussing the general perturbations, let us recapitulate the
consequences of the already studied restricted set of perturbations leading
to the $Z_2\times H_{\nu}^{CP}$ symmetry of the neutrino mass matrix. Three
possible $Z_2$ symmetries  corresponding to  $O$ being a pure rotations in
the ${ij}^{th}$ plane are denoted as $Z_{ij}$. Of these, the $Z_{12}$
symmetry leaves the third column invariant. Thus it can be
phenomenologically consistent as exact symmetry  only when the third column
reproduces  experimental values. The minimum even $n$ for which this
happens
is $n=16$ which gives the third column as
$(0.0253735,0.376842,0.597784)^T.$ 
As already proven in the earlier section, the $Z_{13}$ symmetry
with the  tri-bimaximal mixing at zeroeth order always leads to the
$\mu$-$\tau$ reflection symmetry for 
all the residual symmetry groups labeled as $(n,\gamma,x)=(n,0,x)$. Thus
one needs perturbations which break the $Z_{13}$ symmetry if the original
mixing is tri-bimaximal. The possibility $(n,\gamma,x)=(8,3,x)$ allowed for 
$\Delta(384)$ does not lead to the $\mu$-$\tau$ reflection symmetric
$U_{0}$ and hence also  $U.$ With $Z_{13}$ imposed,
one gets for this case ($s_{23}^2,|\sin\delta|)=(0.43,0.72)$ independent of
$x.$

Implications of the $Z_{23}$ symmetries are quite different. This symmetry
implies  the following correlations between $\cos \delta$ and the 
atmospheric mixing angle:
\be \label{correlation1}
\cos\delta=\frac{\left({c_{23}}^2-s_{23}^2\right)
   \left(c_{12}^2 s_{13}^2-s_{12}^2\right)}{4 c_{12} c_{23}
   s_{12} s_{13} s_{23}} ~.\ee
This correlation is true for the case with tree level TBM mixing. Other
choices of residual symmetries lead to corrections to it which depend on
the angle $\theta_1$ defined in eq.(\ref{thetai}) \cite{Yao:2016zev}.
Eq.(\ref{correlation1}) has been noticed before \cite{Xing:2014zka} and it 
implies negative $\cos\delta$ for $\theta_{23}<\frac{\pi}{4}.$
This is quite consistent with indication of  $s_{23}^2<1/2$ and
$\delta\approx \frac{3\pi}{2}$ at T2K and NO$\nu$A but this relation by
itself cannot fix the quadrant in which $\delta$ lies. This requires the
knowledge of the sign of $\sin\delta$ as well and hence of the Jarlskog
invariant
$J=Im[U_{12}U_{23}U_{13}^*U_{22}^*].$ The sign of $J$ depends on the
ordering of rows of $U_{sym}$ which is not fixed by the symmetry. One can
however derive the following relation
in case of $U_0$ having the TBM form:
\be \label{correlation2}
\frac{J}{c_{13}^2(c_{23}^2-s_{23}^2)}=-\frac{1}{12}\frac{Re[p_1^{-3}]}{Im[p_1
^{-3}]}~.\ee
This relation is invariant under the interchange of the second and the
third row of $U_{sym}.$ Moreover, just like eq.(\ref{correlation1}), this
relation is also independent of the unknown angle $\phi_{23}$ which defines
the $Z_{23}$ symmetry. It then follows that the sign of $J$ is essentially
determined by the
group theoretical factor $p_1^{-3}$  and the quadrant of $2\theta_{23}.$
Eqs.
(\ref{correlation1},\ref{correlation2}) together  serve to fix the quadrant
in which
$\delta$ lies.\footnote{The RHS of eq.(\ref{correlation2}) would change the
sign if the leptonic doublets are assigned to a 3-dimensional
representation conjugate to the one used here since $U_{sym}$ in this case
would go to its conjugate.}

We collect in Table \ref{table1} values of the predicted
$s_{23}^2,\sin\delta$ for various choices of the neutrino residual symmetries. The 
corresponding symmetry $T_l$ for the charged leptons is taken as $Z_3.$ We
have determined $\phi_{23}$ through fits to three mixing angles as
determined in the global analysis of \cite{deSalas:2017kay} choosing the
solution
corresponding to $\theta_{23}<\frac{\pi}{4}.$
It is seen that TBM mixing and  $Z_{23}$ symmetry do not give 
$s_{23}^2$ within
$3\sigma$ at the minimum for most choices of the residual symmetries.
Only exception being $(8,0,3),(8,0,5).$ These two cases lead
to large $|\sin\delta|$ but opposite values of $\sin\delta.$ $\cos\delta$
always remains negative in accordance with the relation
(\ref{correlation1}). 
In the alternative case with $(n,\gamma,x)=(8,3,x)$, the $Z_{23}$ symmetry
can give  $s_{23}^2 $  within 3$\sigma$ for all $x$ of these several $x$
predict relatively large $\sin\delta$ as shown in Table \ref{table1}. 
\begin{table} \begin{center}  \begin{tabular}{|c|c|} \hline \hline 
$(n,\gamma,x)$&($s_{23}^2,|\delta|$)\\ 
\hline
$(2,0,1)$,$(6,0,1)$,$(6,0,5)$&$(0.30,0)$\\ \hline
$(4,0,1)$&$(0.36,227.6^\degree)$\\ 
$(4,0,3)$&$(0.36,132.4^\degree)$\\ \hline
$(8,0,3)$&$(0.42,110.7^\degree)$\\
$(8,0,5)$&$(0.42,249.3^\degree)$\\ \hline 
$(8,0,1)$&$(0.31,155.7^\degree)$\\ 
$(8,0,7)$&$(0.31,204.3^\degree)$\\ \hline
$(8,3,1)$&$(0.48,0^\degree)$\\ \hline
$(8,3,0)$&$(0.46,336^\degree)$\\ 
$(8,3,2)$&$(0.46,24^\degree)$\\ \hline
$(8,3,3)$&$(0.47,346.8^\degree)$\\ 
$(8,3,7)$&$(0.47,13.2^\degree)$\\ \hline
$(8,3,4)$&$(0.48,5.9^\degree)$\\
$(8,3,6)$&$(0.48,354.1^\degree)$\\ \hline
$(8,3,5)$&$(0.43,41.2^\degree)$\\ \hline \hline
\end{tabular} 
\end{center} \caption{Values of $(s_{23}^2,\delta)$ implied by the
best fit solution in case of the $Z_{23}$ symmetry.}
\label{table1}  
\end{table}

We now discuss the effects of adding sizable perturbations to the above
mentioned symmetric limits. One can identify three physically interesting
cases: (A) $\epsilon_{12}+\epsilon_{23}=0$ (B) $\epsilon_{33}=0$
and (C)
$\epsilon_{12}-\epsilon_{23}=0$. Imposing any two of them simultaneously
correspond to imposing various $Z_2\times H_{\nu}^{CP}$ symmetries. Choosing
only one would amount to a single parameter perturbation to these $Z_2$. We
shall first do this exercise in two of the cases (A) and (B). 
Since both the magnitudes and signs of the
unperturbed masses $m_i$ play an important role in determining   values
of the perturbed mixing angles $\phi_{ij}$  defined in eq.(\ref{phiij}),  we take these  masses 
and two of the parameters $\epsilon_{ij}$ as defined in the above 
cases as inputs.
All the three residual symmetries are
still broken in both the cases.  We assume the normal hierarchy  and fit
these parameters to the  results of the global analysis
\cite{deSalas:2017kay} which
includes the latest results from T2K and NO$\nu$A for various  possible
values of $(n,\gamma,x)$.  The CP violating phase is not included in the fit
and thus can be regarded as a prediction. 

\begin{table}[!ht]  \begin{center} 
%\begin{math} 
\begin{tabular}{|c|c|c|} \hline \hline 
$(n,\gamma,x)$&Case A&Case B\\
&$\epsilon_{23}=-\epsilon_{12},\epsilon_{33}\not=0$&$\epsilon_{33}=0$\\
\hline 
(2,0,1)&$(\epsilon_{23},\epsilon_{33})=(0.0217, 0.0999)$&$(\epsilon_{12},
\epsilon_{23})=(-.0999, -0.0388)$\\
&$(m_1,m_2,m_3)=(0.0251, 0.0256, 0.0548)$&$(m_1,m_2,m_3)=(-0.0706,
0.07208, 0.08626)$\\
&$(s_{23}^2,|\sin\delta|,\chi_{min}^2)=(0.455, 0.979,
6.01)$&$(s_{23}^2,|\sin\delta|,
\chi_{min}^2)=(0.301, 0.053, 48.32)$\\ \hline 
(4,0,1)&$(\epsilon_{23},\epsilon_{33})=(-0.0799, 0.0999)$&$(\epsilon_{12},
\epsilon_{23})=(0.0316, -0.0588)$\\
&$(m_1,m_2,m_3)=(-0.0146, 0.0183, 0.0513)$&$(m_1,m_2,m_3)=(-0.054,
0.05588, 0.0727)$\\
&$(s_{23}^2,|\sin\delta|,\chi_{min}^2)=(0.403, 0.905,
2.285)$&$(s_{23}^2,|\sin\delta|, \chi_{min}^2)=(0.357, 0.732, 15.67)$\\
\hline 
(8,0,5)&$(\epsilon_{23},\epsilon_{33})=(0.03316,
0.0447)$&$(\epsilon_{12},\epsilon_{23})=(-0.02754, 0.035498)$\\
&$(m_1,m_2,m_3)=(-0.0626, 0.06414,
0.07954)$&$(m_1,m_2,m_3)=(-0.0707, 0.07223, 0.08585)$\\
&$(s_{23}^2,|\sin\delta|,\chi_{min}^2)=(0.431,0.95,0.032)$&$(s_{23}^2,
|\sin\delta| ,
\chi_{min}^2)=(0.422, 0.935, 0.31)$\\ 
\hline 
(8,3,0)&$(\epsilon_{23},\epsilon_{33})=(-0.0087, -0.0908)$&$(\epsilon_{12}
, \epsilon_{23})=(0.0273,-0.0275)$\\
&$(m_1,m_2,m_3)=(0.009762, 0.01270, 0.05081)$&$(m_1,m_2,m_3)=(0.04424,
0.045394, 0.066827)$\\
&$(s_{23}^2,|\sin\delta|,\chi_{min}^2)=(0.430, 0.693,
0.00015)$&$(s_{23}^2,|\sin\delta|,
\chi_{min}^2)=(.46,0.41,2.86))$\\ \hline 
(8,3,5)&$(\epsilon_{23},\epsilon_{33})=(0.05436,
0.0011603)$&$(\epsilon_{12},
\epsilon_{23})=(0.0484,-0.0561)$\\
&$(m_1,m_2,m_3)=(0.0323, 0.034254, 0.046)$&$(m_1,m_2,m_3)=(0.03776,
0.0377756, 0.0616487)$\\
&$(s_{23}^2,|\sin\delta|,\chi_{min}^2)=(0.426, 0.659,
0.046)$&$(s_{23}^2,|\sin\delta|,
\chi_{min}^2)=(0.426, 0.675, 0.060)$\\ \hline  \hline
\end{tabular} 
%\end{math} 
\end{center} \caption{Results of fits with
two parameter perturbations  to neutrino  residual
symmetries  labeled by $(n,\gamma,x)$ and contained in $\Delta(6 n^2)$
groups for $n=2,4,6,8$. Two separate cases  labeled as $A$ and $B$ are
explained in the text. Table gives input parameters determined from the fits
to neutrino parameters as determined in the global analysis
of \cite{deSalas:2017kay}. The masses are in eV units. We also show the
predicted values of
$s_{23}^2,|\sin\delta|$ at the minimum. 
 }
\label{table2}
\end{table}
%%%%%%%%%%%%%%%%%%%%%

Most of the cases correspond to the TBM mixing since this is
the only possibility which can be cured by small perturbations for $\Delta(6
n^2)$ with $n=2,4,6$. 
As a measure of the smallness of parameters, we have imposed  the
restriction $|\epsilon_{ij}|<0.1$ on  the
perturbation parameters. In addition, we also impose
the constraint that the sum of the fitted neutrino masses should be less
than $0.23$ eV \cite{Ade:2015xua}.
The results of this analysis are summarized in Table \ref{table2}.
The Majorana phase $p_1$ is the only controlling parameter  distinguishing
various symmetries. In case of the TBM, $p_1$ depends only on the ratio
$x/n$, all the symmetries labeled by $(n,0,x)$ and $(m n,0,m x)$ with
integer $m$
lead to the same result and we have listed the lowest member in cases
presented in the table. 
 We do not display all the cases, but present only
some specific interesting examples. It is found that in most of the cases,
the obtained minimum displays either approximate $Z_{13}$ or $Z_{23}$
symmetry although neither was imposed to start with. The sub-dominant
contributions however play  important roles in
improving fits as discussed below. We first discuss case (A).\\ \\
\noindent {\bf Case A} 
\begin{itemize}
\item  $(n,\gamma,x)=(2,0,1)$ is  the only possibility within
the smallest group
$S_4$ which does not give the  exact $\mu$-$\tau$ reflection symmetry for
arbitrary values of $\epsilon_{ij}.$ One gets vanishing $\delta$ in this
case if $Z_{23}$ symmetry is exact and maximal for the exact $Z_{13}$
symmetry. When both are broken and $\epsilon_{33}$ and $\epsilon_{23}$ are
present one gets departures from the exact $\mu$-$\tau$ reflection symmetry
and quite large $|\sin \delta|.$ This can be attributed to
more dominant  $\epsilon_{33}$ compared to $\epsilon_{23}$ at the minimum.
The latter however leads to the required departures from the maximal value
of $\theta_{23}.$  
\item The next case is the  group $\Delta(96)$ with $(n,\gamma,x)=(4,0,x).$
The cases $(n,\gamma,x)=(4,0,1),(4,0,3)$ give identical $s_{23}^2$ while
$(n,\gamma,x)=(4,0,0),(4,0,4)$ give the exact $\mu$-$\tau$ reflection
symmetry. The result for the $(4,0,1)$ is displayed in the
Table \ref{table2}. The exact $Z_{23}$ symmetry with $\epsilon_{33}=0$ does
not give $s_{23}^2$ within the 3$\sigma$ and exact $Z_{13}$ predicts the
$\mu$-$\tau$ reflection symmetry. 
This changes when both $\epsilon_{33}$ and $\epsilon_{23}$ are present. Now
both  have comparable values at the minimum 
and the resulting value of $s_{23}^2$ differs from
the
exact $Z_{23}$  or $Z_{13}$ symmetry.
\item The next group in the series is $\Delta(216)$ and different  cases
are distinguished by  values of $x$ in $(n,\gamma,x)=(6,0,x)$. The cases
with $x=0,2,4,6$ all give the exact $\mu$-$\tau$ reflection symmetry for any
values of $\epsilon_{ij}$ as already argued. The remaining two cases
$x=1$ and $x=3$ give identical $s_{23}^2.$ These case are identical to the results of $(2,0,1)$ and  are not displayed.
\item The next group is $\Delta(384)$ with $n=8.$ This allows two
possibilities namely, $\gamma=0$ and $3$ both of 
which give quite good zeroeth order result. For this group, 
all the cases with $(n,\gamma,x)=(8,0,x)$ for $x=1-6$  give very good fit
and predict large $|\sin\delta|.$  The case with  $x=7$ 
on the other hand lead to a good fit but predict relatively small
$|\sin\delta|.$, The cases $(8,0,3)$ and $(8,0,5)$ give better fit
than others. In these cases, neither $\epsilon_{33}$ nor $\epsilon_{23}$
dominates
and they have comparable values. 

The case with  $\gamma=3$ differs from the previous ones since the leading
order mixing is not TBM and thus 
one always gets departures from the exact $\mu$-$\tau$ symmetric limit. One
gets $s_{13}^2\sim 0.011$
and $s_{23}^3\sim 0.41$ at the leading order in this case. The obtained
minimum in the presence of perturbations $\epsilon_{33},\epsilon_{23}$ 
displays nearly   $Z_{13}$
symmetry for all $x$ and result for a specific cases $(8,3,0),(8,3,5)$ are
shown in
the table. The obtained values are quite close to the $Z_{13}$ symmetric 
limit $s_{23}^2\sim 0.426,|\sin\delta|\sim  0.72.$
\end{itemize}
\noindent{\bf Case B}
\begin{itemize}
\item Unlike the case (A),  addition of the parameter $\epsilon_{33}$ to the
$Z_{23}$ symmetric case 
does not change results compared to the $Z_{23}$
symmetric case and the values of $(s_{23}^2,|\sin\delta|)$ are close to the
ones displayed in Table \ref{table1}. 
Thus only,  $(8,0,5),(8,0,3)$  case for the
TBM give results with correct $s_{23}^2$. 
The case $(4,0,1)$ also displayed in table however
comes close to predicting $s_{23}^2$ with 3$\sigma$. It also
gives large $\sin\delta.$   
\item All the case $(n,\gamma,x)=(8,3,x)$ not having TBM at the zeroeth order can
fit the angles very well and all the solutions display approximate
$Z_{23}$ symmetry. But only the cases $(8,3,0),(8,3,2)$ and $(8,3,5)$ give
large $|\sin\delta|$ respectively, $0.41$ and $0.67$. Solutions for 
$(8,3,0),(8,3,5)$ are displayed in Table \ref{table2}.
\end{itemize}

We have taken only two of the three parameters as non-zero in
the numerical fits presented above. 
It is important to consider the most general case with all the three
parameters present and ask how far the above predictions remain true in the
presence of the third parameter. Rather than fitting global $\chi^2$, we
carried out a
general analysis of this case by randomly varying all three parameters
$\epsilon_{ij}$ in the range
$-0.2-0.2$. The lowest mass mass $m_1$ is varied in the range $(-0.1\sim
0.1)$
eV. The other two masses are chosen positive and $\leq 0.1$ eV. We worked
out predictions for the CP phase in this
situation by demanding that angles as well as the solar and atmospheric
scales lie within their $3\sigma$ range as determined in
\cite{deSalas:2017kay}. This is done for all possible
symmetry choices $(n,0,x)$ (TBM) with $n=2,4,6,8.$ 

\begin{figure}[!ht]
\centering
\includegraphics[width=0.5\textwidth]{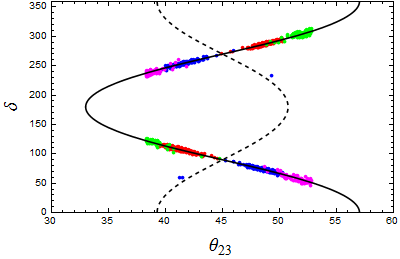}
\caption{ Correlation between  $\theta_{23}$ and $\delta$ obtained in the
presence of the breaking of the residual $Z_2\times Z_2$ symmetry contained
in the $\Delta(6 n^2)$ groups for $n=4$ and $8.$ The considered residual
symmetries are labeled by $(n,0,x)$ and correspond to the TBM mixing
pattern in the absence of perturbations. All the obtained points reproduce
the mixing angles within 3$\sigma$ as determined in \cite{deSalas:2017kay}.
Various symmetries are labeled as $(4,0,1)$ (magenta),$(4,0,3)$ (geen), $(8,0,3)$
(red) and $(8,0,5)$ (blue). Curves  obtained
assuming exact
$Z_{23}\times H_\nu^{CP}$ (continuous) and $Z_{13}\times H_\nu^{CP}$
(dashed)  symmetry are also shown for comparison.}
\label{fig1}
\end{figure}

\begin{figure}[!ht]
\centering
\includegraphics[width=0.5\textwidth]{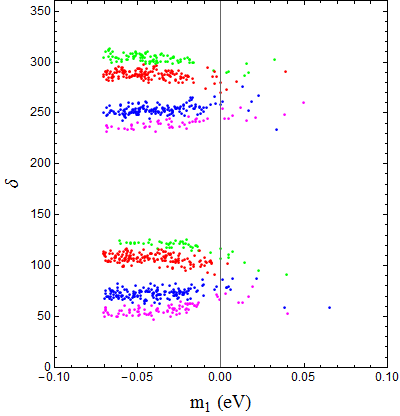}
\caption{Plot of the lowest unperturbed neutrino mass $m_1$ versus
$\delta$ for  the residual symmetries labeled by  
$(n,\gamma,x)=(4,0,1)$ (magenta) and $(4,0,3)$ (green), $(8,0,3)$(red)  and $(8,0,5)$(blue). The allowed points
reproduce all the three mixing angles in the  3$\sigma$ range as determined
in global fits \cite{deSalas:2017kay}.}
\label{fig2}
\end{figure}

The numerical analysis of the cases (A) and (B)  shows two
patterns. There exists several residual symmetries, e.g. symmetries labelled by
$(2,0,1)$ and $(8,3,0)$ in\footnote{Other examples of such symmetries are $(6,0,1),(6,0,5),(8,0,1),(8,0,7)$, $(8,3,1),(8,3,4).$} Table \ref{table2} 
for which the
predicted $\delta$ at the minimum are quite different in two cases (A) and
(B). In contrast symmetries $(4,0,1),(4,0,3),(8,0,3),(8,0,5),(8,3,5)$
predict similar values of $\delta.$ It would be expected that the
predictions of $\delta$ would lie in a narrow range in these cases  when all
the three
parameters are present. This is indeed the case and we present predictions
of the cases which lead to TBM mixing in the absence of perturbations. One
finds definite correlations 
 between  $\theta_{23}$ and $\delta$ and these  are displayed
in Fig.\ref{fig1} for four
specific choices of $(n,0,x).$ 
For comparison, we also show the curves obtained in case of the $Z_{23}$
and $Z_{13}$ symmetry assuming best fit values for
$\theta_{12},\theta_{13}$ using the analytic expression as given in 
eqs.(\ref{correlation1}) for the $Z_{23}$ symmetry (continuous curve)  and
similar one obtained assuming $Z_{13}$ symmetry (dotted curve).
\vspace*{1cm}
\vspace*{0.5cm}
Remarkably, all the allowed points obtained these cases are clustered around
the $Z_{23}$ symmetry curve even though this symmetry  is not assumed to start with.
Moreover, we find that the allowed values of $|\epsilon_{ij}|$ are
comparable
in large number of cases 
and are not restricted to the exact $Z_{23}$ symmetric solution
$\epsilon_{12}=-\epsilon_{23},\epsilon_{33}=0$ considered earlier. In spite
of this, one seems to be getting an effective $Z_{23}$ symmetry. The reason
can be understood from Fig.\ref{fig2} which displays  variation of
$\delta$ with the lowest mass $m_1$ for the same choices of $(n,0,x)$ as in
the case of Fig.\ref{fig1}. It is seen that most of the solutions correspond
to quasi-degenerate spectrum and occur when $m_1$ is negative relative to
$m_{2,3}$ which are assumed positive in the analysis. This results in
effective suppression of $\phi_{12},\phi_{13}$ compared to $\phi_{23}$ as 
seen from approximate expressions given in eq.(\ref{phiij}). This
results in effective $Z_{23}.$ As explicitly seen in Fig.\ref{fig2}, the
predicted range of $\delta$
is characteristic of the underlying symmetries rather than the values of
$\epsilon_{ij}.$
%\vspace*{1cm}
%\vspace*{0.5cm}
The allowed points in Fig.\ref{fig1} are of two types. For the symmetries
$(4,0,3),(8,0,5)$ one gets $\delta$ in the third (second)  quadrant if
$\theta_{23}<1/2 (\theta_{23}>1/2$). Reverse situation arises for the other
symmetries $(4,0,1),(8,0,5)$. This is quite consistent with the expression
eqs.(\ref{correlation1},\ref{correlation2}) and reflect the approximate
$Z_{23}$
symmetry. 

Before we end the section, we give an explicit example from the points
obtained in our random analysis. The chosen example 
corresponds to the neutrino residual symmetry $(n,\gamma,x)=(8,0,5)$
contained in the $\Delta(384)$ group.The following values of the input
parameter in this case are found to reproduce all the angles within
3$\sigma$ and lead to large $\sin\delta$
\beqa \label{803case}
(\epsilon_{33},\epsilon_{23},\epsilon_{12})&=&
(-0.030842, 0.0264886, -0.0741432)~,\nonumber \\
(m_{1},m_2,m_3)&=& (-0.0533103, 0.0555739, 0.0718049)~ {\rm eV} .\eeqa
These values lead to the $U_{PMNS}$ matrix
\be \label{upmnsexample}
U_{PMNS}=\left(
\begin{array}{ccc}
 0.827613\, +0.00472979 i & -0.499131-0.212111 i &
   -0.130806-0.0616621 i \\
 0.192143\, +0.347111 i & -0.0955472+0.653941 i & 0.613431\,
   -0.171843 i \\
 0.18395\, -0.351841 i & 0.271839\, -0.44183 i & 0.720233\,
   +0.233505 i \\
\end{array}
\right)~.\ee
This leads to 
\be \label{angles}
(s_{13}^2,s_{23}^2,s_{12}^2,\delta)=(0.0209, 0.4145,
0.3004,247.9^\degree)~\ee
Interchanging the second and the third row of eq.(\ref{upmnsexample})
results in a solution with $s_{23}^2=0.585$ and $\delta=67.9^\degree$ 
The
perturbatively generated matrix $O$ in eq.(\ref{upmns}) is given in this
case by
\be \label{o}
O=\left(
\begin{array}{ccc}
 0.99963 & 0.0158604 & 0.0221184 \\
 0.0214073 & -0.960027 & -0.279087 \\
 -0.0168078 & -0.279457 & 0.960011 \\
\end{array}
\right)
~.\ee
The role of the relative signs of the unperturbed masses is clear from this.
The angle $\phi_{23}$ gets considerably enhanced compared to the basic
parameter $\epsilon_{-}$ determining it and other angles are relatively
suppressed. As a result, $O$ correspond approximately to a rotation in the $2-3$ plane
reflecting the $Z_{23}$ symmetry.

\section{Summary}
Flavour symmetries are widely used for understanding the observed patterns
of neutrino mixing angles and phases. A very predictive theoretical
hypothesis of  invariance of $M_\nu$ ($M_lM_l^\dagger$) under residual
$Z_2\times Z_2$ ($Z_n$) symmetry leads to a prediction of the vanishing
Dirac CP phase if these residual symmetries are embedded in the $\Delta(6
n^2)$ groups. The neutrino Majorana phases can also be predicted by
extending the neutrino symmetry to $Z_2\times Z_2\times H_\nu^{CP}$
\cite{King:2014rwa}. An alternative approach known as semi-direct approach
assumes the residual symmetry $Z_2\times H_\nu^{C}$ with $H_\nu^{CP}$
commuting with the $Z_2$
\cite{Feruglio:2012cw,Holthausen:2012dk,Chen:2014wxa,Li:2016ppt,
Hagedorn:2014wha,Ding:2014ora,Yao:2016zev}. This can predict  non-zero
$\delta.$ 

We have presented  here a simple, straightforward and predictive
generalization of above schemes and explored it both analytically and
numerically.  This generalization is suitable to investigate the effects of
perturbations to 
the original $Z_2\times Z_2$ symmetry. These perturbations are assumed to respect the 
GenCP consistent with the original $Z_2\times Z_2$ symmetry.
As shown in section (2), one gets a simple 
expression for the $U_{PMNS}$ mixing matrix involving three unknown mixing
angles and  group theoretically determined parameters  $p_1$ or $p_2$ and $\theta$ in this
case.
Using this, it is shown that $Z_2\times Z_2$ symmetries leading to TBM mixing
invariably lead to maximal Dirac phase if CP symmetry is unbroken in the
symmetric case and Majorana phases assume specific values.
This interesting result is essentially due to $\mu$-$\tau$
reflection symmetry which always arises in the said circumstances as an
effective symmetry of the neutrino mass matrix in the flavour basis in all
the relevant $\Delta(6 n^2)$ groups.

The cases in which the unperturbed mixing matrix contain non-trivial Majorana phases
are studied numerically for the $\Delta(6 n^2)$ groups for $n=2,4,,6,8.$ and
the predicted 
Dirac CP phases are worked out. As discussed in detail, 
there exist several residual symmetries for which the predicted CP phases
are characteristic of the symmetry rather than the values of the
perturbation parameters as long as these parameters  are required to
reproduce the
other mixing angles correctly. One finds very definite correlations between
the quadrants in which 2$\theta_{23}$ and $\delta$ lie as displayed in
Fig.\ref{fig1}. These are explicitly worked out for the symmetries labeled
as $(4,0,1),(4,0,3),(8,0,3),(8,0,5)$ all of which lead to the TBM mixing in
the absence of perturbations. Interestingly, one finds a presence of
underlying approximate $Z_{23}\times H_\nu^{CP}$ symmetry in these cases
even  when  perturbations  significantly  break this
symmetry. Similar correlations are expected to  exist in case of other
residual symmetries which do not give TBM, e.g. the symmetries labeled as
$(8,3,x)$ in $\Delta(384).$ These are not studied but can be explored using
the present formalism. The present study was restricted to explore the
consequences of the said assumptions from the symmetry considerations
rather than building  specific models.
\section{Acknowledgements} I thank  Ketan M. Patel for a careful
reading of the  manuscript and helpful suggestions.
This work was supported by BRNS (Department of Atomic Energy) and by
Department of Science and
Technology, Government of India through the Raja Ramanna  fellowship and the
J. C. Bose grant respectively.
\bibliography{pertref.bib}
\bibliographystyle{apsrev4-1}
\end{document}